\documentclass[journal, 12pt, draftclsnofoot, onecolumn]{IEEEtran}
\makeatletter
\def\ps@headings{%
\def\@oddhead{\mbox{}\scriptsize\rightmark \hfil \thepage}%
\def\@evenhead{\scriptsize\thepage \hfil \leftmark\mbox{}}%
\def\@oddfoot{}%
\def\@evenfoot{}}
\makeatother
\usepackage{subfigure}
\usepackage{graphicx}          % Include this line if your
\usepackage{cite}
\usepackage{url}
\usepackage{amssymb,amsmath}

\usepackage{blindtext}

\begin{document}
\title{Optimizing TCP Performance in Multi-AP Residential Broadband Connections via Mini-Slot Access}

\author{Domenico Giustiniano$^*$, Eduard Goma$^\#$, Alberto Lopez Toledo$^\#$,\\ George Athanasiou$^+$ 
\\$^*$ETH Zurich, Switzerland\\$^\#$Telefonica Research,
Barcelona, Spain\\$^+$KTH - Royal Institute of Technology, Stockholm, Sweden}

\maketitle

\begin{abstract}
The high bandwidth demand of Internet applications has recently driven the need of increasing the residential download speed. A practical solution to the problem has been proposed aggregating the bandwidth of 802.11 Access Points (APs) backhauls in range via 802.11 connections. Since 802.11 devices are usually single-radio, the communication to multiple APs on different radio-channels requires the introduction of a time-division multiple access (TDMA) policy at the client station.
Current investigation in this area supposes that there is a sufficient number of TCP flows to saturate the Asymmetric Digital Subscriber Line (ADSL) behind the APs. However, this may be not guaranteed according to the user traffic pattern.
As a consequence, a TDMA policy introduces additional delays in the end-to-end transmissions that will cause degradation of the TCP throughput and an under-utilization of the AP backhauls. In this paper, we first perform an in-depth experimental analysis with a customized 802.11 driver of how the usage of 
multi-AP TDMA affects the observed Round-Trip-Time (RTT) of TCP flows. Then, we introduce a simple analytical model that accurately predicts the TCP RTT when accessing the wireless medium with a Multi-AP TDMA policy. Based on this model, we propose a resource allocation 
algorithm that runs locally at the station and it greatly reduces the observed TCP RTT with a very low computational cost. Our proposed scheme can improve up to $1.5$ times the aggregate throughput observed by the station compared to state-of-the-art multi-AP TDMA allocations. We also show that the throughput performance of the algorithm is very close to the theoretical upper-bound in key simulation scenarios. 
\end{abstract}

\section{Introduction}

Asymmetric digital subscriber line (ADSL) has become the `de-facto' standard for residential broadband access to the Internet. In addition, the density of ADSL deployments with 802.11 Wireless Local Area Network (WLAN) connectivity tends to be high, specially in urban areas~\cite{han}.
The interplay between these two technologies introduces interesting technical challenges and opportunities that can be exploited. First, WLAN access rates are typically an order of magnitude higher than the bottleneck of the end-to-end path, which is either the ADSL~\cite{siekkinen} or the backbone~\cite{maier}. Second, the set of ADSL links in the neighborhood are generally under-utilized~\cite{siekkinen}. As a consequence, there is potential to bundle the Access Points (APs) backhaul bandwidth via 802.11 connections. However i) APs usually operate on independent radio-channel and ii) users typically connect to these APs with single-radio commodity 802.11 cards. 
%EG: Changed the end of the paragraph. There was a too long sentence difficult to follow.
Because a single-radio card cannot simultaneously connect to more than one AP, it has been proposed to rely on the standard 802.11 Power Saving (PS) mode to implement a Time-Division Multiple Access (TDMA) policy.
Stations spend enough time to either collect all the bandwidth from each AP~\cite{fatvap} or to provide a fair sharing of the aggregated resources~\cite{themis} by sequentially cycling through the APs in a round-robin fashion~\cite{virtualwifi}.

\begin{figure}
\centering
\includegraphics[height=7cm]{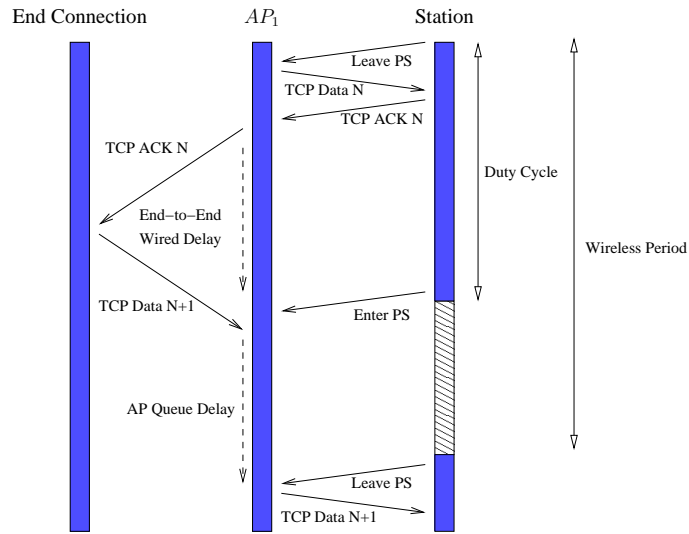}
\caption{Relation between TCP congestion control and time-division access to multiple APs.}
\label{fig:RTO}
\end{figure}

Unfortunately, multi-AP TDMA policies hurt the throughput performance of single TCP flows by increasing their Round-Trip-Time (RTT). To illustrate this effect, we consider the scenario in Fig.~\ref{fig:RTO}, where one station is connected to $N$ APs. We focus on the time that the station spends connected to one of the APs, say $\mathrm{AP}_1$. 
%We denote the \emph{wireless period} as the total amount of time to cycle through all the APs, and the \mathrm{duty cycle} as the percentage of time the station spends on $\mathrm{AP}_1$. 
%At the beginning of the duty cycle, 
When the station is connected to $\mathrm{AP}_1$, it starts receiving the buffered TCP data packets. While connected, the station normally receives TCP data and replies with TCP ACKs. These TCP ACKs will trigger the transmission of new TCP data from the sender. Because of the end-to-end wired delay, these TCP data may arrive to $\mathrm{AP}_1$ right after the station has already moved to the next AP. These packets will be buffered by $\mathrm{AP}_1$ until the station connects
again to it. The result is that the time duration of the connectivity and non-connectivity periods may result in TCP flows observing a RTT artificially larger than the actual end-to-end wired delay.

Motivated by the aforementioned problem, the main contributions of this paper are:
\begin{enumerate}
\item an in-depth analysis of the effect of multi-AP TDMA on TCP flows by performing numerous experiments using our prototype station that connects to multiple APs. 
\item an analytical model that correlates the TCP congestion control with the multi-AP TDMA policy.
\item a cost-efficient resource allocation algorithm, named \emph{min-max disconnection time}, that distributes the time spent by the station to the APs in minislots (part of the TDMA slot) to minimize the total disconnection time from the APs. 
\end{enumerate}

We evaluate our model and show that it accurately fits the experimental results. We further study the algorithm via extensive simulations and we show that it performs very close to the theoretical upper bound. The rest of the paper is organized as follows. Section~\ref{sec:related_work} reviews related work,
and Section~\ref{sec:TDMA_to_multipleAPs} presents our prototype implementation of multi-AP TDMA used in the experimental tests. Then, Section~\ref{sec:model_TCP} investigates the performance degradation of TCP on the multi-AP TDMA scenario, and it introduces an accurate analytical model. Section~\ref{sec:validation} validates the analytical model via both experiments and simulations. The resource allocation algorithm is 
introduced and validated in Section~\ref{sec:resource_allocation}. Finally Section~\ref{sec:conclusions} concludes the paper.

\section{Related work}
\label{sec:related_work}
The need for 802.11 resource allocation schemes has been extensively studied in the literature~\cite{Athanasiou07, Tassiulas, Athanasiou09}. Many of the proposed schemes rely on either non-standard compliant features~\cite{multimac}, or completely develop an entire new MAC protocol~\cite{freemac}. Both strategies may be undesirable, and so we avoid them. Given that, the resource allocation scheme that more closely relates to ours is~\cite{oml}, that studied the problem of absence of application-specific 802.11 resource allocation schemes. As a solution, they designed
and implemented an overlay MAC layer (OML) to divide the time into slots of equal size. Then, they used a distributed algorithm to allocate the slots across the competing nodes, where each competing node receives a number of slots proportional to its weight function.
However, the authors let as an open issue the understanding of the increased delay for TCP flows in presence of the slotted mechanism~\cite{oml}.

Although overlay solutions are easy to be implemented, they are often sub-optimal and difficult to scale because of the overlapping and duplication of similar functionalities at different layers (e.g. in the driver and in the card firmware). The VirtualWiFi project~\cite{virtualwifi} proposed an architecture that abstracts a single 802.11 WLAN card to appear as multiple virtual clients to the user. Each client instance adopts standard PHY/MAC protocols, but it can be separately configured at the driver level. An interesting application was the idea of connecting to multiple APs through a single radio interface.
The authors rely on the 802.11 Power Save (PS) mode feature to switch among different 802.11 WLAN nodes in a time-division fashion. A station can inform the current 802.11 WLAN node that it is going into PS mode --- so that it can buffer packets directed to it --- and switch the radio-frequency to other 802.11 WLAN nodes, only to come back to the original node before the PS period expires. 

Based on the above PS mechanism, FatVAP~\cite{fatvap} studied the problem of ADSL bandwidth aggregation via wireless connectivity. The authors introduced a local scheduler that computes the percentage of time to dedicate to each AP in order to maximize the aggregate throughput at each client station. The solution leverages on the fact that the high speed wireless card at the station needs to be connected to each AP for a short period of time in order to collect all the pending data.  
THEMIS~\cite{themis} reformulated the problem considering that gross unfairness would be generated in the above setting. Their approach achieved weighted proportional fairness and they experimentally validated it with a multistory building and real ADSLs, showing that it outperformed previous solutions. However, both FatVAP and THEMIS do not explore TCP latency-related problems for single TCP flows. They essentially limited the analysis to scenarios with sufficient number of TCP flows, such that the ADSL bandwidth is saturated, or with short-lived TCP flows, where the congestion control phase does not trigger. Finally, Juggler~\cite{juggler} proposed an architecture similar to one in~\cite{fatvap} and it focused on the support of a seamless hand-off between WLAN APs.

\section{Connecting to Multiple APs with Off-The-Shelf Hardware}
\label{sec:TDMA_to_multipleAPs}

In this section, we briefly describe WiSwitcher~\cite{presto09}, the experimental 802.11 station that we have implemented and it will be
used in the rest of this paper as the basis for the experimental tests.

\begin{figure}[t]
\centering
\subfigure[Topology.]{\includegraphics[height=5cm]{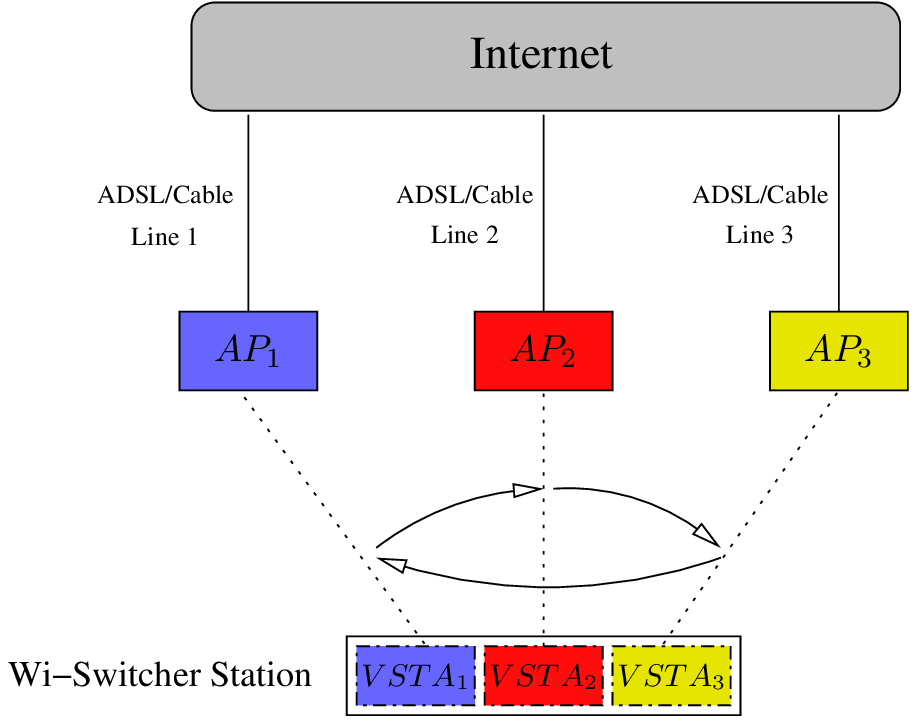}
\label{fig:ClubADSL_topology}}
\hfil
\subfigure[Relation between duty cycle and wireless period.]{\includegraphics[height=2cm]{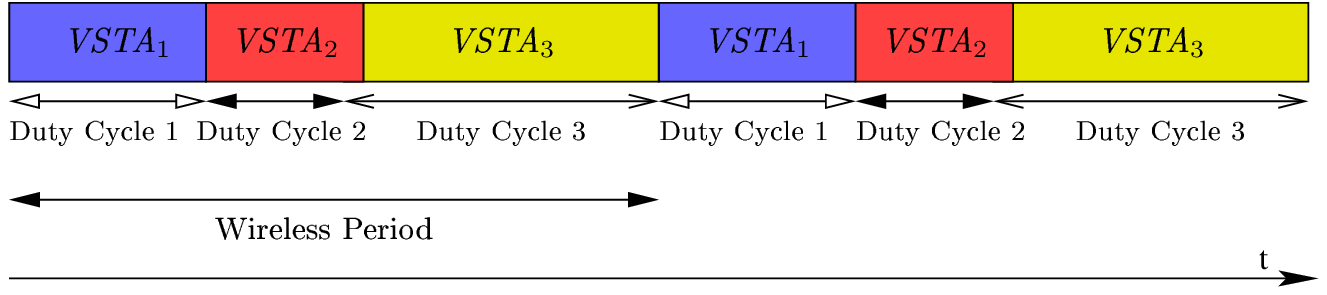}
\label{fig:duty_wp}}
\caption{Time Division Access to Multiple APs}
\label{fig:xxx}
\end{figure}

In WiSwitcher, the wireless driver of the single radio interface is \emph{virtualized}, i.e., it appears as independent Virtual STAtions ($\mathrm{VSTA}_i$) associated to their respective Access Points $\mathrm{AP}_i$. Each $\mathrm{VSTA}_i$ connects to Internet via $\mathrm{AP}_i$, regardless of the $\mathrm{AP}_i$ radio-frequency. Let us consider the scenario with one station in Fig.~\ref{fig:ClubADSL_topology}, connected to three APs. In this example, WiSwitcher configures $3$ virtual stations $\mathrm{VSTA}_1$, $\mathrm{VSTA}_2$ and $\mathrm{VSTA}_3$. Each of these VSTAs connects to one ADSL via its AP in range.

As we can see in Fig.~\ref{fig:duty_wp}, WiSwitcher assigns the control of the card to a $\mathrm{VSTA}_i$ for a given percentage of time, called \emph{duty cycle} $f_i$ (with $\sum_i f_i = 1$).
During this time, $\mathrm{VSTA}_i$ transmits and receives frames using the $\mathrm{AP}_i$ backhaul, while the other VSTAs (and the corresponding APs) can only buffer packets. We denote \emph{wireless period} $T$ as the amount of time to cycle through all the VSTAs.
A summary of the main variables used in this section and the rest of the paper is given in Table~\ref{tab:variables}.

\subsection{MAC Protocol}
WiSwitcher manages the multiple backhaul connections relying on the 802.11 PS mechanism.
Particularly, referring to the example in Fig~\ref{fig:duty_wp}:

\begin{enumerate}
\item During the reserved duty cycle, $\mathrm{VSTA}_1$ transmits and receives data according to the 802.11 Distributed Coordination Function (DCF) protocol.
The other VSTAs are dormant in PS mode, and hence they (and the corresponding APs) can only buffer packets.
\item When the duty cycle expires, $\mathrm{VSTA}_1$ sends a frame to inform $\mathrm{AP}_1$ that is going to PS mode and waits for its MAC ACK.
According to the 802.11 protocol, $\mathrm{AP}_1$ starts to buffer the packets directed to it.
\item WiSwitcher assigns the control of the card to $\mathrm{VSTA}_2$ and switches to the $\mathrm{AP}_2$ radio-frequency.
\item $\mathrm{VSTA}_2$ sends a frame to announce that it can send and receive traffic and it waits for its MAC ACK.
\item The process continues until the station has cycled through all the $\mathrm{VSTAs}$  (a wireless period $T$).
\end{enumerate}

\begin{table}[t]
\centering
\begin{tabular}{|c|c|} \hline
$\mathrm{AP}_i$ & $i$-th AP \\ \hline
$N$ & Number of AP backhauls  \\ \hline
$T~~(ms) $ & Wireless period  \\ \hline
$\mathrm{VSTA}_i$ & $i$-th Virtual STAtion, associated to $\mathrm{AP}_i$ \\ \hline
$d_i$ & End-to-end wired delay\\ \hline
$f_i~~(\leq 1)$ & Duty cycle for the $i$-th virtual station \\ \hline
$g_i~~(\geq 1)$ & Number of slots for the $i$-th virtual station \\ \hline
$G$ & Total number of slots \\ \hline
$C_{j}$ & Disconnection cost for the $j$-th slot \\ \hline
$\mathrm{SlotTime}$ & Minimum slot size\\ \hline
\end{tabular}
\caption{Main variables used.}
\label{tab:variables}
\end{table}

In the implementation, we incur in a channel-switching cost --- i.e. the time where WiSwitcher cannot send and receive any traffic --- of $1.2$\,ms for uplink traffic and $1.5$\,ms for downlink traffic. This cost is less than half of the one obtained in the time-division implementation given in~\cite{fatvap, juggler}. This result has been achieved using 802.11 standard-compliant solutions, such as a MAC virtual queue per AP, and an efficient management of a hardware buffer size of one (1) data packet. The bulk of the cost of WiSwitcher is caused by the hardware operation delay, which is in the order of $800$\,$\mu$sec in our Atheros
chipset-based cards. This cost is hardware dependent and in other chipset implementations is reduced to $200-500$\,$\mu$sec~\cite{juggler}.

Furthermore, the implementation achieves a fine-grained timing at MAC/PHY level, thus avoiding any additional variance of the delay in the packet transmission, and it considers an independent instance of the rate selection algorithm for each VSTA. This allows us to connect to APs with different link qualities. The reader is referred to~\cite{presto09} for an in-depth description of the MAC implementation.

\subsection{Network Layer Functionalities}
At the network layer, three functionalities are needed:

\textbf{Scheduler}: it calculates the percentage of time (duty cycle) to spend on each AP in order to maximize some utility function.
In this work, the duty cycles are fixed via user-space commands.

\textbf{Load balancer}: it assigns the new TCP flow from the upper layers to the different VSTAs so that the total load received from each AP maintains the proportions indicated by the scheduler. Since the load balancer is not the main objective of this work, we use the same per-flow basis scheduler presented in~\cite{fatvap}. 

\textbf{Reverse-NAT}: In order to guarantee transparency to higher layers, we implement a reverse-network address translation (NAT) module with two functions: i) ensure that the packets leave the host with the correct source IP address (i.e. the one corresponding to the outgoing VSTA, as assigned by the AP) and ii) that the incoming packets are presented to the OS with the expected IP address, a dummy IP address in our implementation. Reverse-NAT modules were also present in~\cite{fatvap} and~\cite{juggler}.

\begin{figure}[t]
\centering
\includegraphics[height=6cm]{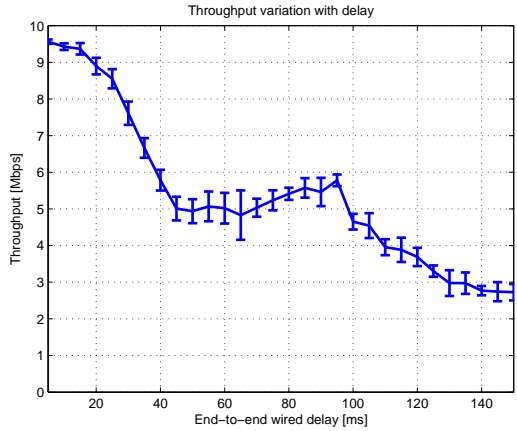}
\caption{Experimental throughput connected 50\% of time to one AP}
\label{fig:50percentage_thput}
\end{figure}

\section{TCP over Multi-APs TDMA}
\label{sec:model_TCP}

In this section, we first show an experimental test that enlightens the correlation between the end-to-end throughput and the delay added by the TDMA policy,
and then we introduce an analytical model that characterizes the TCP RTT for a station connected to multiple APs.
The importance of the model is that it not only gives insights into the problem, but it will be also used to validate the resource allocation algorithm later defined in the paper.

\subsection{Example of TCP Throughput over Multi-AP TDMA policy}
We experimentally test the side-effect of the multi-AP TDMA policy on a (long-lived) TCP session.
Figure~\ref{fig:50percentage_thput} shows the average TCP throughput obtained by a station spending $50\%$ of its time (that is duty cycle of $f_1=0.5$) to one AP,
as a function of the end-to-end wired delay $d_1$. For the test, we consider a wireless period $T=100$\,ms, which gives a connection of $f_1 T=50$\,ms on $\mathrm{AP}_1$. Each point is the average TCP throughput obtained over $5$ independent tests of $300$\,s. In the tests, the average experimental congestion signal rate we measured is of $\approx 0.4$\%. These losses
are likely generated on the wireless link\footnote{As a result of 
using SACK, TCP congestion signals are mainly caused by fast retransmissions due to duplicated ACKs because
its goal is to avoid retransmission timeouts. For all SACK-based TCPs, multiple losses within one RTT are treated as a single congestion signal.
In this paper we use congestion signal and packet loss interchangeably, and always refer to losses at TCP layer.}.

%Since the RTT is a function of the disconnection time from the AP $(1-f_1) T$ and the end-to-end wired delay $d_1$, 
We can see from Fig.~\ref{fig:50percentage_thput} that the station gets a similar throughput for both $d_1=50$\,ms and $d_1=100$\,ms. This is caused by the similar RTT observed for $50$ and $100$\,ms. The result is that the TCP data arriving at the AP with $50$\,ms of end-to-end wired delay have to wait for an extra-buffering time at the AP, due to the disconnection of $(1-f_1) T=50$\,ms. This reduces the throughput observed by the TCP flow for $d_1=50$\,ms. More exactly, there are small valleys in the throughput: a disconnection of $50$\,ms increases the buffer size of the AP. Further increasing the end-to-end wired delay from $50$ to $100$\,ms, there is a higher and higher probability that downlink packets will arrive at the AP when the station is connected to it. This will reduce the buffering time at the AP (and thus the observed RTT), that will deliver the packets in a shorter time, with a slight increase of the throughput\footnote{Note that there are slight variations in the packet losses observed in the experimental tests, which translate in variations (represented by error-bars in the figure) of the average throughput observed between different experiments.}.

\subsection{Modeling the TCP RTT over Multi-AP TDMA}
\label{subsec:model_delay}
\begin{figure}[t]
\centering
\includegraphics[height=4.8cm]{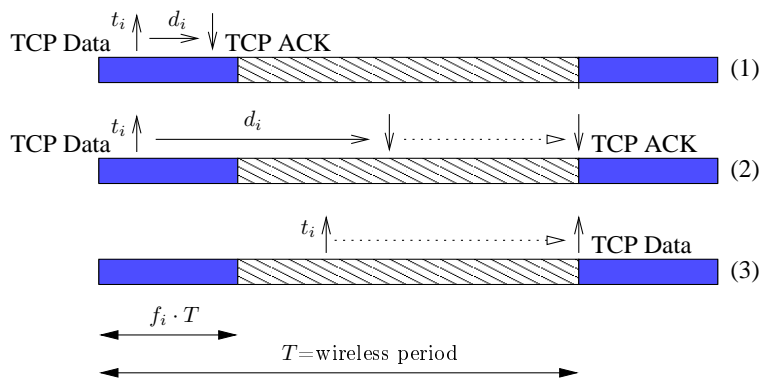}
\caption{Model of the relation between TCP congestion control and duty cycle}
\label{fig:model}
\end{figure}

We can model the dependency of the TCP RTT on the end-to-end delay and the duty cycle by observing all the possible cases in which TDMA affects the observed RTT. In what follows, we consider the uplink case
(e.g. the VSTA is sending data to a remote server), but it is straightforward to see that the RTT computations are symmetric for
both the uplink and downlink cases. We distinguish three conditions:
\begin{enumerate}
\item We consider the case of Fig.~\ref{fig:model}(1) in which the station sends the TCP data at time $t_i$, during its duty cycle. Also we assume that the end-to-end delay $d_i$ is such that the TCP ACK arrives from the TCP server before the station disconnects from the AP.
%, i.e., $t_i + d < f_i T$.
In that case, we see that the observed RTT from TCP is $d_i$.
\item Next, we consider the case of Fig.~\ref{fig:model}(2), where the station sends the TCP data at time $t_i$,
during its active period, but the end-to-end delay $d_i$ is such that the TCP ACK arrives from the server during
the time reserved to other VSTAs. The $AP_i$ will buffer the packet in its queue
until the station reconnects again at time $0$ of the next wireless period. In this case, the observed RTT for
the TCP packets is $ T - t_i $, where $T$ is the wireless period. Note that, as long as $(1 - f_i)T > (d_i \mod T)$, there will be always
some packet that will wait in the $\mathrm{AP}_2$ downlink buffer because of the disconnection period.
\item Finally, we consider the case of Fig.~\ref{fig:model}(3), wherein the TCP data is buffered at the station at time $t_i$, during the sleeping period in the AP. However, we experimentally verified by monitoring the AP queues that case (3) does not occur in the TCP steady state, and no new TCP data is buffered during the sleeping period. The reason is that, in the TCP steady state case, new TCP data can only be sent when a TCP ACK is received from the server. But as we have seen, the TCP ACKs can only arrive to the station when it is connected to the AP.
\end{enumerate}

Finally, in order to take into account that the station takes some time for processing and transmitting the TCP ACKs, as verified experimentally, we consider that:
i) the TCP ACKs arrive exponentially distributed over the duty cycle, because there is a high probability that some TCP ACK 
is waiting in the $\mathrm{AP}_i$ downlink buffer when the station re-connects to it (in the beginning of a duty cycle) ii) one TCP DATA
is sent right after the reception of a TCP ACK. This assumption considers the case when, in average and steady state, throughput does not
either increase or decrease
iii) When $f_i T$ is very small, some of the TCP data scheduled during the connection period will inevitably be sent at the next
connectivity period, due to buffering delay. Based on these assumptions, we statistically calculate the RTT distribution, considering as input the
wireless period $T$, the duty cycle $f_i$ and the end-to-end delay $d_i$. Despite the simplicity of the model, Section~\ref{subsec:tcprttmodelvalidation} will show that the model matches the experimental results.

\subsubsection*{Mapping the modeled TCP RTT to Throughput}
\label{sub:mathis}
Although a wide variety of TCP algorithms are used on the
Internet, the current most popular implementation is TCP Reno~\cite{reno}.
Then, in order to map the RTT given by the model to throughput, we use the Mathis 
TCP model~\cite{mathis}, which is intended to predict TCP end-to-end throughput as: $BW \leq \frac{\mathrm{MSS}}{\mathrm{RTT}} \cdot \frac{1}{\sqrt{\mathrm{p}}}$, where $\mathrm{RTT}$ is the Round-Trip-Time observed by the station, $\mathrm{MSS}$ is the TCP Maximum Segment Size and $\mathrm{p}$ is the packet loss
rate\footnote{This model applies to long lived connections over nearly all implementations of TCP Reno with SACK TCP.
Note that, in order to use this model, the packet loss rate should be smaller than $2$\%, 
condition verified in all the experimental tests in this paper.}.

\section{Evaluation}
\label{sec:validation}

In this section:
\begin{itemize}
\item we validate the accuracy of the TCP RTT model presented in the previous section, comparing it with experimental results.
\item we show that long disconnection time severely affects the TCP throughput.
\item we demonstrate that, for any duty cycle, the best strategy is to reduce the wireless period $T$ as much as possible.
\item we show that the selection of the wireless period $T$ must be done based on the smallest duty cycle $\min_i f_i$ of the station. 
\end{itemize}
In what follows we discuss the details of the experimental and simulation setup.

\textbf{Experimental setup}
In each controlled test, we use laptops with Atheros-based chipsets running WiSwitcher as described in Section~\ref{sec:TDMA_to_multipleAPs} and off-the-shelf APs (Linksys) with DD-WRT v24sp1 firmware. On the wireless station, automatic rate selection, wireless multimedia extensions, and the RTS/CTS mechanism are disabled. In the experimental tests of this paper, the 802.11 PHY rate is fixed to $54$\,Mbps. Other tests were performed in other configurations (e.g. with automatic rate selection enabled), and showed similar results to the ones presented in this paper.
Any no-802.11 standard compliant features at the MAC level were also disabled. Our station use an hardware queue with best effort parameters.

For the transport layer, we use a Linux standard TCP Reno with SACK and delayed ACK options enabled. TCP parameters are monitored using a modified version of the TCP probe kernel module and the kernel patch Web100.
For each test, we establish one TCP connection over an AP backhaul\footnote{Note that the link utilization can increase establishing more than one TCP connection over each AP\cite{presto09}, which is out-of-the-scope of this paper.} and we collect statistics using the \emph{iperf} tool. Regarding the wired connections, we emulate the AP backhaul links through the \emph{tc} Linux traffic shaper, varying the delay using the \emph{netem} tool.

For each experimental test, we establish one TCP Reno connection over each AP, ran $5$ independent tests of $300$\,secs and plot the average values obtained. To achieve independent tests, the station is configured so that the TCP metrics are reset after each test.

\textbf{Simulation setup}
The simulations are performed using the model described in Sec.~\ref{subsec:model_delay}, implemented in MATLAB using as input the experimental values of
MSS and TCP congestion signals rate per packet for the Mathis formula defined in Section~\ref{sub:mathis}~\cite{mathis}.

\begin{figure*}[t]
\centerline{
\subfigure[Experimental RTT]{\includegraphics[height=6cm,width=8cm]{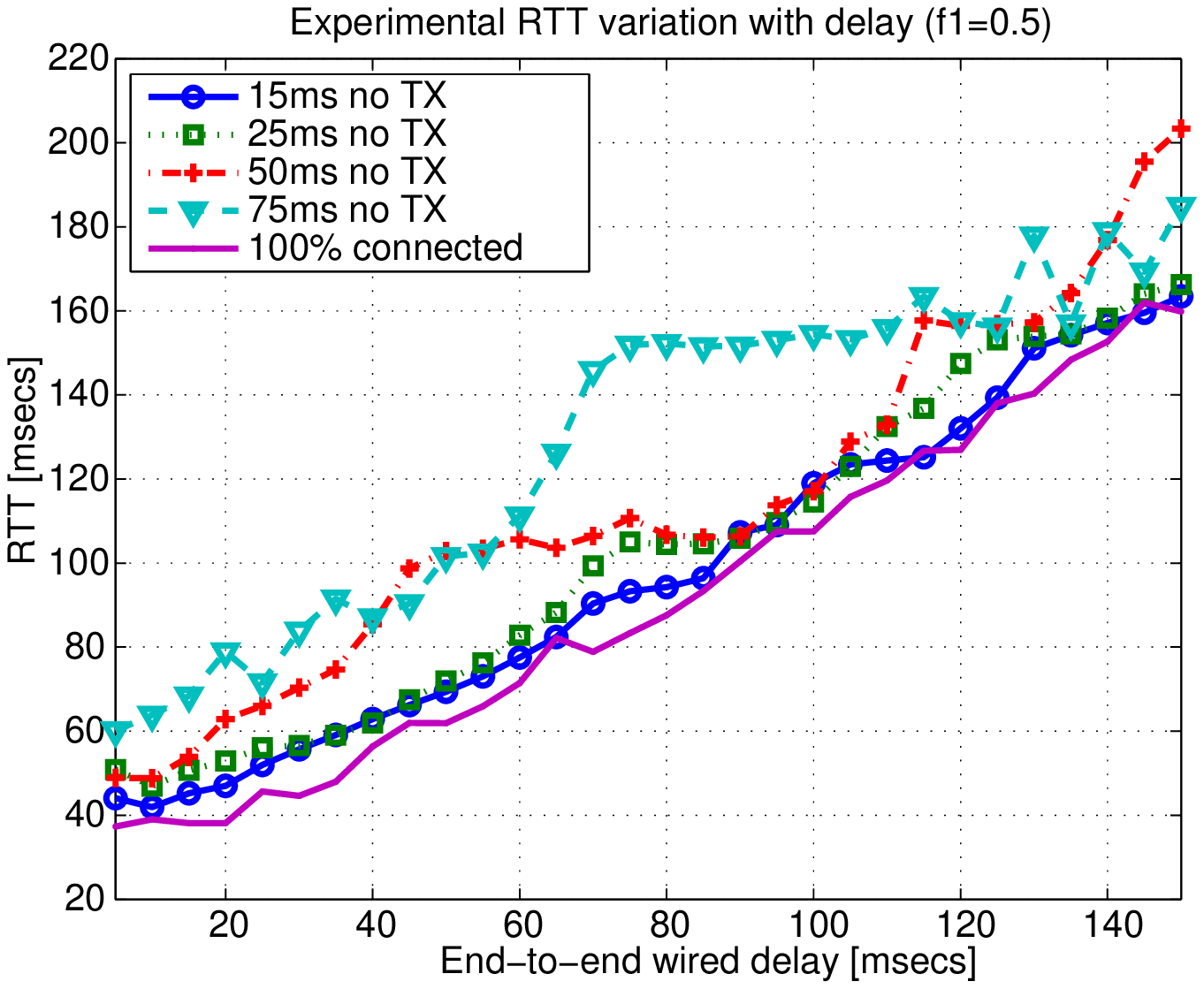}
\label{fig:rtt50real}}
\hfill
\subfigure[Analytical RTT.]{\includegraphics[height=6cm,width=7.5cm]{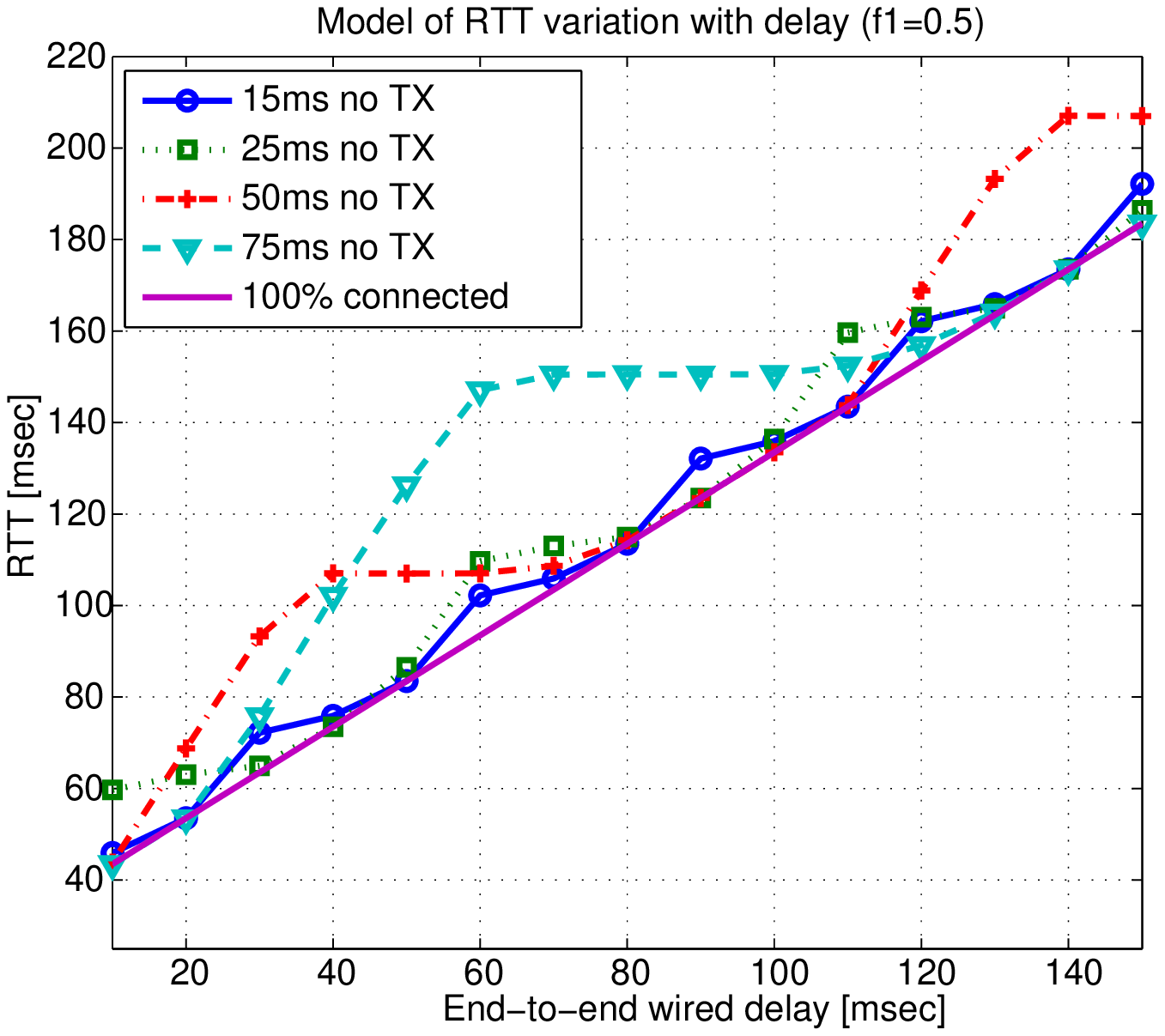}
\label{fig:rtt50model}}}
\caption{Downlink RTT for $f= 0.5$ to one AP. The station is not connected to the AP for a time equal to $0$, $15$, $25$, $50$ and $75$\,ms, respectively. We see that the analytical model accurately predicts the measured TCP RTT.}
\label{fig:rtt_f05}
\end{figure*}

\subsection{TCP RTT Model Validation}
\label{subsec:tcprttmodelvalidation}

In this section we compare the RTT values achieved experimentally with the ones of the model. For brevity, we only show one scenario, but similar finding have been achieved with several other setting (e.g.
with different values of duty cycle).

Fig.~\ref{fig:rtt50real} shows the average RTT values obtained experimentally as a function of the end-to-end wired delay, when the WiSwitcher station spends $50\%$ of its time connected to an AP, and it observes a disconnection time of $0$, $15$, $25$, $50$ and $75$\,ms, respectively. The plot shows that the increase of disconnection time significantly affects the measured TCP RTT. 

Fig.~\ref{fig:rtt50model} shows the observed TCP RTT calculated using the model for the same scenario. We can see that the analytical model accurately predicts the measured TCP RTT.
The observed differences are the result of the variable losses observed in the experiments and the expected noise in the experimental environment. 

Furthermore, there are specific situations where a smaller disconnection time results in a higher RTT. As an example, we consider the RTT in Fig.~\ref{fig:rtt50model}, for a disconnection time of $50$ and $75$\,ms and an end-to-end wired delay of $125$\,ms. Here, the interplay of the disconnection time and the delay causes a higher TCP RTT for a disconnection time of $50$ ms compared to the $75$\,ms case.
This phenomenon gets less important at smaller disconnection times (see e.g. the RTT observed with a disconnection time of $15$ and $25$\,ms).

\begin{figure*}[t]
\centering
\centerline{
\subfigure[Throughput with $50$\% of connection ($f_1=0.5$) to $AP_1$.]{\includegraphics[width=8.5cm]{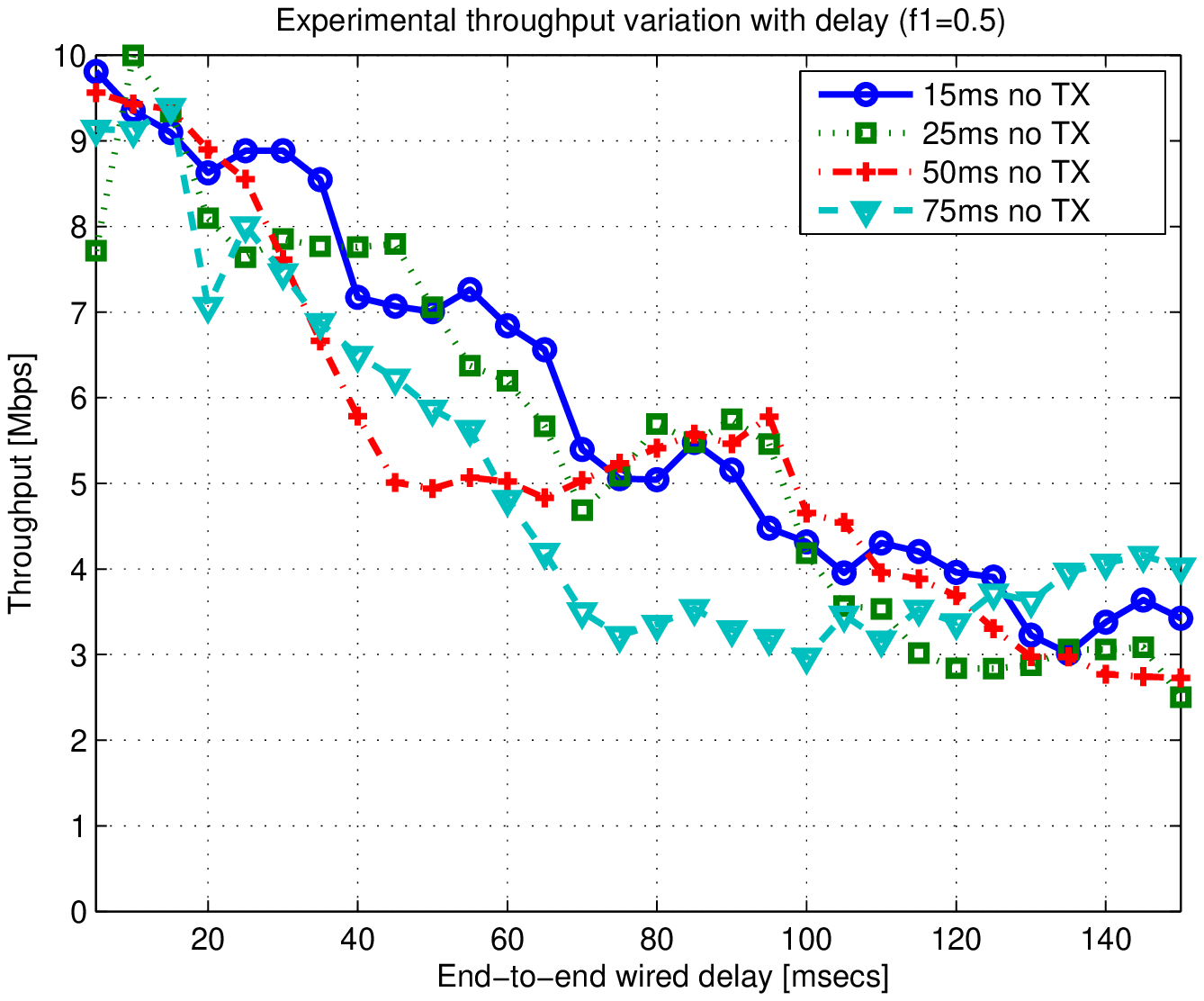}
\label{fig:throughput50real}}
\hfil
\subfigure[Throughput with $10$\% of connection ($f_1=0.1$) to $AP_1$.]{\includegraphics[height=6.3cm]{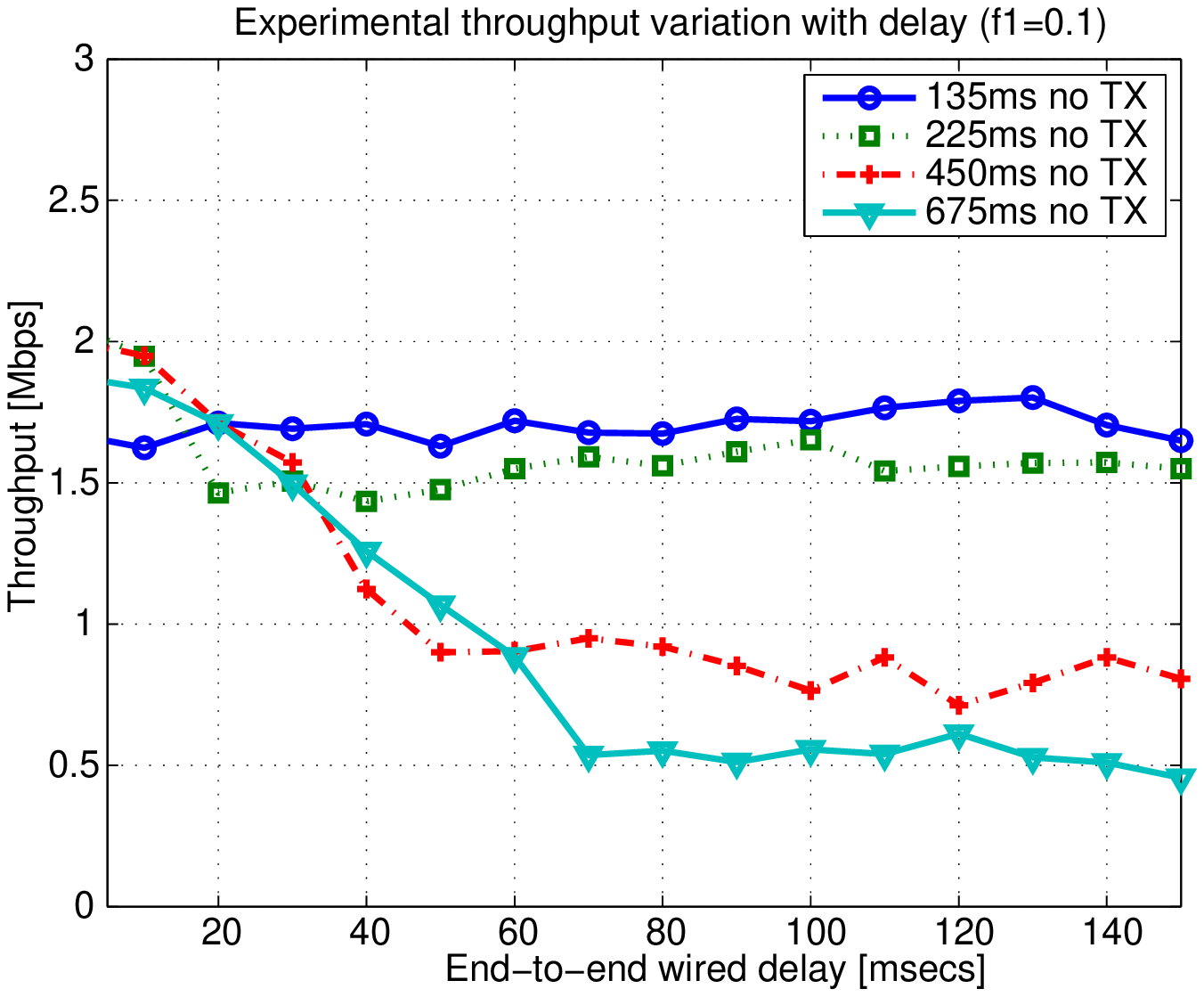}
\label{fig:tput_f01real}}}
\caption{Throughput per TCP-flow with different duty cycles and disconnection times}
\label{fig:xxx}
\end{figure*}

\subsection{Impact on Throughput per TCP flow}

In this section we study the throughput observed by a TCP flow opened on one AP.
Fig.~\ref{fig:throughput50real} depicts the experimental throughput when the \emph{duty cycle} of one AP is $50$\%. We observe that even for
small delays, the throughput performance may be dramatically affected.
As an example, when we operate with a disconnection time of $75$ ms, we observe a quasi-constant throughput when the end-to-end wired delay
spans from $75$ to $150$ ms. This is caused by the similar RTT observed at $75$ and $150$ ms of end-to-end wired delay.

Even more evident is the case where the VSTA is connected for the same amount of time --- hence for a connection time of $f_1 T$ = $\{15, 25, 50, 75 \}$ ms --- \emph{but} for a smaller \emph{duty cycle} to $\mathrm{AP}_1$, e.g. $10$\% of its time. We can also see from Fig.~\ref{fig:tput_f01real} that the penalty in throughput is more severe as the disconnection time grows. For example, when the disconnection time is 675 ms, the average throughput is more than three times smaller than the throughput achieved when the disconnection time is $135$ or $225$\,ms.

\subsection{Reducing the Wireless Period}
\label{sub:wireless_period}

\begin{figure}[t]
\centering
\includegraphics[height=6.5cm]{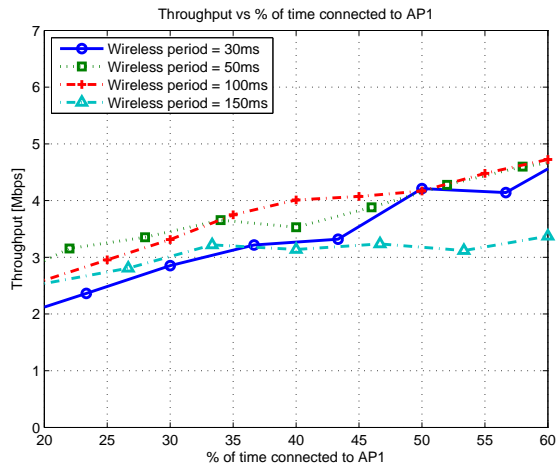}
\caption{Experimental downlink throughput connected 50 \% of time to one AP for an end-to-end delay of 100 ms}
\label{fig:min_wireless_period}
\end{figure}

Based on the analysis in the previous section, in order to reduce the impact of TDMA on the TCP throughput, we have to keep the disconnection time as small as possible. Since the disconnection time is equal to $T-f_iT$=$(1-f_i)T$, this also implies that, for a fixed $f_i$, the wireless period $T$ should be kept small. We study this issue with experimental tests. Figure~\ref{fig:min_wireless_period} shows the throughput achieved as a function of the percentage of time connected to one AP. In the tests, we use different wireless periods ($T$ = $\{30, 50, 100, 150\}$ ms) and we fix the end-to-end delay to $100$\,ms. The figure show that similar performance are achieved with wireless periods of $T=50$\,ms and $T=100$\,ms, while throughput can be severely affected choosing a wireless period of $T=150$\,ms, once the time of connection is above $45\%$. 

However, $T$ cannot be reduced as much as we want. For $T=30$\,ms, the station gets slightly less throughput for small \emph{duty cycles}. In fact, there are two limiting factors: i) there is a time spent by the 802.11 card to switch AP (called \emph{switching cost}, as described in details in Section~\ref{sub:slotted_operation}) ii) the frequent AP switching introduces an extra congestion signal rate of $\approx 0.01-0.08$\%, caused by an inefficient management of the transmission queues at the driver in off-the-shelf APs\cite{presto09}. As a practical design aspect, since the congestion signal rate and the switching cost affect more severely smaller duty cycles, we
can conclude that the selection of the wireless period $T$ should be based on the smallest duty cycle at the station. In the next section, we provide an accurate description of the selection of the wireless period $T$ and its correlation with the other parameters.

%Once we studied the performance observed on a given AP, in the next section we face the problem of optimizing the performance observed by the station in \emph{all} the APs it is connected to, by defining and validating a resource allocation algorithm with a low computational cost.

\section{Increasing the Aggregated Throughput}
\label{sec:resource_allocation}

In this section we aim at improving the aggregate throughput at the station by 
i) introducing the assignment of slots to each VSTA, and ii) allocating the slots via a distributed resource allocation algorithm. The objective is to minimize the total disconnection time such as the TCP throughput of single flows is increased. 

\subsection{Concept of Slotted Operation}
\label{sub:slotted_operation}
Instead of connecting to each $\mathrm{AP}_i$ for a consecutive amount of time $f_i \cdot T$, we introduce the concept of slot assignment and we give $g_i \geq 1 $ slots to each $\mathrm{VSTA}_i$.
For this scope, according to the analysis in Section~\ref{sub:wireless_period}, we first define $\mathrm{SlotTime}$ as the minimum amount of time allowed in the system at
which the effect of the switching cost and the packet losses can be neglected on the connection with the smallest duty cycle.

In order to select such a $\mathrm{SlotTime}$, based on our empirical data, we get at least $85$\% of the expected throughput --- defined as the throughput
that would be achieved without any cost of switching --- with only $6$\,ms of connection time over a wireless period of $12$\,ms, at least $90$\%
with $10$\,ms over $20$\,ms, and at least $95$\% of the expected throughput with a connection time of $15$ ms over a wireless period of $30$\,ms (see~\cite{presto09} for
details). These values are lower-bounds, because achieved when the AP queue is constantly backlogged. With Internet traffic, APs are backlogged only in the beginning of the duty cycle (because
of the downlink packets already in the AP queue at the time of starting the duty cycle),
while instead transmit at the end-to-end transmission rate for the rest of the duty cycle~\cite{fatvap}. Then, $\mathrm{SlotTime} = 10-15$\,ms safely gets the expected throughput.

However, an allocation of slots of equal length $\mathrm{SlotTime}$ would increase the wireless period $T$,
with immediate drawbacks on the performance. As an example, if the station is connected to two APs
with scheduler output $f_1=0.27$ and $f_2=0.73$, we would need $27$ slots for $\mathrm{AP}_1$ and $73$ slots on $\mathrm{AP}_2$. A $\mathrm{SlotTime}$ set to $15$ ms would result in a wireless period of $15\cdot 100=1500$\,ms, which is computational inefficient.

Driven by the experiments and simulations of the previous section, we resolve this problem with the following principles:
\begin{itemize}
\item we calculate the wireless period as: $T= \frac{\mathrm{SlotTime}}{\min_i f_i}$, that is, the procedure reduces the wireless period $T$, based on the smallest duty cycle of the station (as demonstrated in Section~\ref{sub:wireless_period}).
\item we derive the number of slots locally assigned to each $\mathrm{VSTA}_i$ as: $g_i=\lfloor (f_i T)/\mathrm{SlotTime} \rfloor$, for a total number of slots of $G=\sum g_i$.
\item we determine the slot size per $\mathrm{VSTA}_i$ as: $\mathrm{SlotTime}_i = \frac{f_i T}{g_i}$. This may give slots of different sizes
among different $\mathrm{VSTA}s$. 
\end{itemize}
Note that the solution can be transparently applied to the systems proposed in~\cite{themis, fatvap, juggler}, considering the different switching costs of these systems to compute $\mathrm{SlotTime}$.

Once selected $T$, $\{g_i\}$ and $\{\mathrm{SlotTime}_i\}$, our objective is to construct a resource allocation algorithm that, given the set of duty cycles $f_i$ provided by the upper-layer scheduler, it assigns the set of slots to
the $\mathrm{APs}$ in order to minimize the overall disconnection time for all the APs.
For a rigorous analysis of the resource allocation algorithms, some definition is needed, as introduced in the next paragraph.

\textbf{Disconnection Cost} Let us define $\mathbf S_i=[S_i(1), S_i(2), \ldots S_i(g_i)]$ the vector that indicates the slot positions in
the range $[1,G]$ for $\mathrm{VSTA}_i$, with $S_i(g_i+1)=S_i(1)$ and $S_i(l) \neq S_j(m)$ for any $i,j=1,\ldots N$, with $i\neq j$, $l=1,2,\ldots g_i$ and $m=1,2,\ldots g_j$. Besides, we define the cost (slot duration) of each slot as the slot size of the $\mathrm{VSTA}_i$ that uses the slot:
\[C_{S_i(l)}=\mathrm{SlotTime}_i~~~\forall i.\]
%where $i=1,\ldots N$ and $l=1,2,\ldots g_i$.

In order to measure the disconnection cost of the $\mathrm{VSTA}_i$ during two transmissions in the slots $S_i(l)$ and $S_i(l+1)$ we take into account the costs of the intermediate slots $C_{S_i(l)+1},\ldots C_{S_i(l+1)-1}$. Therefore, we introduce the following cost function:
\[{c_{i,l}} = \sum\limits_{j = {S_i}(l) + 1}^{{S_i}(l + 1) - 1} {C{}_j}~~~l=1,2,\ldots g_i \]

\emph{Example:} Let us suppose that $N$=$3$ and that the slots are allocated as follows: 
$[\mathrm{VSTA}_1~\mathrm{VSTA}_2\\~\mathrm{VSTA}_3~\mathrm{VSTA}_1~\mathrm{VSTA}_2~\mathrm{VSTA}_1]$.
This gives: $S_1=[1~4~6]$, $S_2=[2~5]$ and $S_3=[3]$.
Furthermore, we suppose that $\mathrm{SlotTime}_3$ = $10$ ms, $\mathrm{SlotTime}_1$ = $12$ ms and $\mathrm{SlotTime}_2$ = $15$ ms.
Then, we calculate the disconnection cost between $S_1(1)$=$1$ and $S_1(2)$=$4$ as $c_{1,1}= C_2 + C_3 = 15 + 10 = 25$ ms.

\subsection{Resource allocation algorithm}
\label{subsec:raa}

We now present three different, fully decentralized, slot allocation mechanisms with different performance and computational costs that aim to reduce the impact that the multi-AP TDMA has on single TCP flows\footnote{We also tested
an allocation mechanism with random assignment of the slots, using as a constraint that each slot is assigned to a given $\mathrm{VSTA}_i$
with a probability equal to $f_i$.
Although this random assignment may decrease the buffering time at the AP in certain configurations,
we found that it generally increases the jitter observed by TCP, and then reduces the observed downlink throughput.}.

\textbf{Blind Resource Allocation}. 
We have seen in Section~\ref{subsec:tcprttmodelvalidation} that a multi-AP TDMA policy increases the observed RTT of the TCP packets. We have also seen that this increase is exactly the disconnection time in the worst case. In other terms, for $\mathrm{VSTA}_i$, and given an allocation that produces a disconnection time of $ \mathop {\max} \limits_{l=1,2,\ldots g_i}{c_{i,l}}$, we would have
\[\mathrm{RTT}_i = d_i + \mathop {\max} \limits_{l=1,2,\ldots g_i}{c_{i,l}}.\]
The TCP throughput
achieved by the above allocation can be approximated as
\[\frac{\mathrm{MSS}}{[d_i + \mathop {\max} \limits_l{c_{i,l}}]\cdot \sqrt{\mathrm{p_i}}},\]
where $\mathrm{MSS}$ and $\mathrm{p_i}$ are the parameters of the Mathis model defined in Section~\ref{sub:mathis}~\cite{mathis}.
It follows that, in order to minimize the throughput penalty caused by disconnection, we need to solve the following problem:
\begin{eqnarray}
\min\sum\limits_i^{N}  \left( \frac{\mathrm{MSS}}{d_i\cdot \sqrt{\mathrm{p_i}}} - \frac{\mathrm{MSS}}{[d_i + \mathop {\max} \limits_l{c_{i,l}}]\cdot \sqrt{\mathrm{p_i}}} \right).
\label{eq:upper_bound}
\end{eqnarray}

The slot assignment obtained from solving (\ref{eq:upper_bound}) depends on the correct estimation of the loss rates $\{p_i\}$ and end-to-end delays $\{d_i\}$. 
In a realistic deployment, an accurate prediction of these values may be not available. In the absence of any end-to-end delay information we can reformulate the problem simply as the minimization of the inverse of the maximum disconnection times as follows:
\begin{eqnarray}
\begin{array}{l}
 \mathop {\max }\limits_{{S_i}(l)} \sum\limits_i^{N} {1/({\mathop {\max }\limits_l {c_{i,l}}}})  \\\\
 s.t.~~\sum\limits_{i = 1}^G {{C_i} = T}  \\
 ~~~~~~{f_i}T = g_i\cdot{\mathrm{SlotTime}_i}~~~~\forall i\\\
 ~~~~~{S_i}(l) \in \{ 1,~G\}~~~~~\forall i,l,\\
\
 \end{array}
\label{eq:brute_force}
\end{eqnarray}
where the variables $C_i$, $\mathrm{SlotTime}$, $f_i$ and $g_i$ are defined in Table~\ref{tab:variables}.

\textbf{Min-Max Disconnection Time Allocation Algorithm}. The blind resource allocation algorithm defined above can be prohibitively expensive. In order to reduce its complexity, we define a \emph{min-max disconnection time} heuristic approach.
This algorithm considers that, in average,
the TCP throughput is more severely affected by the
amount of time that each $\mathrm{VSTA}_i$ is \emph{not connected} to the corresponding $\mathrm{AP}_i$. Therefore, the algorithm tries to minimize the disconnection time, starting with the connection with the largest duty cycle (i.e., the AP backhaul with the highest utilization). 
The algorithm operates as follows:
\begin{enumerate}
\item First, it allocates the slots to the $\mathrm{VSTA}$ with $\max(g_i)$.
\item Next, the $\mathrm{VSTA}s$ with lower number of slots will be served one by one.
At each step, the selected $i$-th $\mathrm{VSTA}_i$ analyzes \emph{only} the slots not previously assigned and
it calculates the vector $\mathbf S_i$ to satisfy the condition: ${\min} {\max\limits_{l=1,2,\ldots g_i}}~(S_{i}(l+1)-S_{i}(l))$,
that is, it selects the $g_i$ slots to minimize the maximum distance between each pair of consecutive slots
assigned to the $\mathrm{VSTA}_i$.
\item Finally, at the last step, the remaining set of slots are assigned to the $\mathrm{VSTA}$ with $\min(g_i)$. The last $\mathrm{VSTA}_i$ to allocate is the one with the smallest duty cycle. For that one, $\mathrm{SlotTime}$ is already chosen such as its performance is not affected
\end{enumerate}

\textbf{Upper-bound}
We also calculate the upper-bound for the TCP aggregate throughput:  
for each delay, we compute the TCP aggregate throughput for all the feasible solutions and select
the one that achieves the maximum throughput. Note that this upper bound can not be calculated in practice and we
include it for comparison purposes.

\subsection{Simulation results}

We now evaluate the above algorithms --- and particularly the aggregate throughput achieved by the \emph{min-max disconnection time}
allocation --- via simulations in different scenarios: i) high number of APs ii) VSTAs with different duty cycle and delay and
iii) different slot size per VSTA. 
In the tests, we consider one long-lived TCP flow for each VSTA and we suppose that
the end-to-end communication is limited by the RTT delay, so that we do not reach the maximum capacity of the end-to-end path. For each test,
we generate $10000$ samples of RTT\footnote{Similar results were achieved with $1000$ samples. More samples then $10000$ can be used,
but it would adversely affect the simulation time, without impact on the results.} and we use a congestion signal rate of $0.32$\%, as the one measured experimentally by
connecting to one AP in range with high signal-to-noise ratio, and measuring the average number of congestion signals per acknowledged 
packet\footnote{We use the congestion signal rate connecting to one AP, because, as discussed in Section~\ref{sub:wireless_period}, current
off-the-shelf APs implementations add at each AP a certain packet loss rate, that may limit the performance in experimental implementations.
The reader is referred to~\cite{presto09} for more details. In this work, we do not consider these implementation issues, (that,
however, do not currently allow a validation of the resource allocation algorithm with real experiments), and assume that the limiting
factor is the time spent by the 802.11 card to switch AP.}. Note that each TCP flow experiences a different RTT, according to the
specific duty cycle and slots' assignment. Note also that more TCP flows may be sent per AP and other congestion signal rates may be used
(even higher, according to the link quality and the 802.11 PHY rate), but do not add a new contribution to this section.

\begin{figure}[t]
\centering
\subfigure[Slot distribution]{\includegraphics[height=5.5cm]{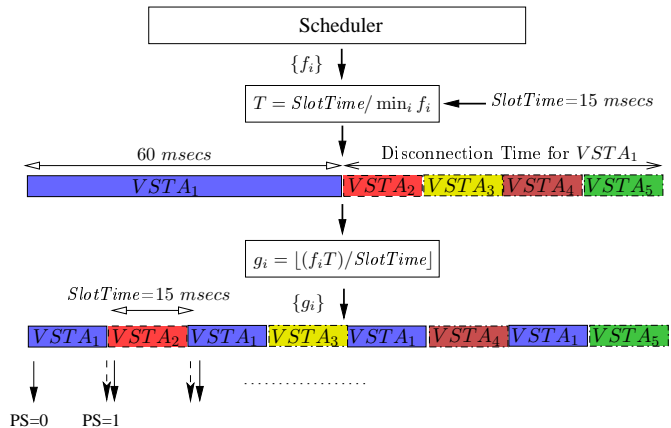}
\label{fig:MAC_TDMA_old}}
\hfil
\subfigure[Throughput improvement.]{\includegraphics[height=5.8cm]{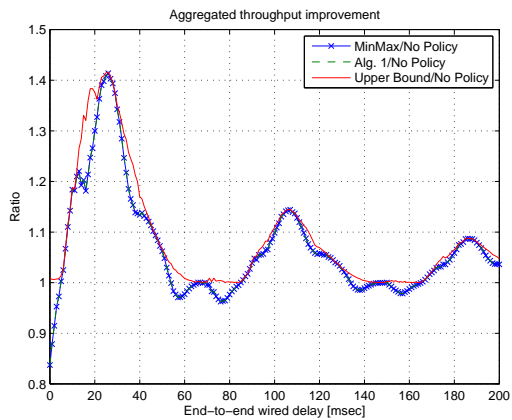}
\label{fig:throughput_with_Min-Distance_case1}}
\caption{Min-max disconnection time allocation algorithm. Case 1.}
\label{fig:xxx}
\end{figure}

\textbf{Case 1: High number of APs}.
Fig.~\ref{fig:MAC_TDMA_old} shows a station is connected to $5$ APs, with the scheduler selecting the following duty cycles: $f_1=0.5$, $f_2=0.125$, $f_3=0.125$, $f_4=0.125$, $f_5=0.125$.
The corresponding number of slots, based on the slot calculation presented in Section VI.A, are $g_1=4$, $g_2=1$, $g_3=1$, $g_4=1$, $g_5=1$, and
the total number of slots is $G=8$.
In the model, we use $\mathrm{SlotTime}= 15$\,ms, which gives a wireless
period of $15$\,ms $\cdot 8$\,slots $= 120$\,ms.
The algorithm minimizes the time without transmission allocating first the slot to $\mathrm{VSTA}_1$, then to $\mathrm{VSTA}_2$,
$\mathrm{VSTA}_3$,  $\mathrm{VSTA}_4$, and $\mathrm{VSTA}_5$.

Fig. \ref{fig:throughput_with_Min-Distance_case1} depicts the throughput improvement
versus the end-to-end delay, obtained comparing the proposed allocation algorithm (labeled ``MinMax'') with no resource allocation (labeled ``No Policy''), that is, spending consecutive 60 ms on $\mathrm{VSTA}_1$, and then $15$\,ms on $\mathrm{VSTA}_2$, $\mathrm{VSTA}_3$, $\mathrm{VSTA}_4$, and $\mathrm{VSTA}_5$, sequentially. 

We observe that the \emph{min-max disconnection time} allocation improves the throughput in all the cases, thanks to the reduction of the disconnection time. The min-max algorithm improves the throughput by up to $1.5$ times respect to the case without any resource allocation. Note that for an end-to-end wired delay of $0-5$ ms, the \emph{min-max disconnection time} algorithm has a slight lower aggregate throughput. This is because, with this very small delay, the higher number of AP switching increases the probability that the TCP packet needs to wait at the next connection period before being ACKed.

Fig.~\ref{fig:throughput_with_Min-Distance_case1} also depicts the throughput achieved by running the algorithm in (\ref{eq:brute_force}) (labeled ``Alg. 1''). We can see that the heuristic approach performs identically to the blind resource allocation. 

Finally, we run a test with all possible slot allocations. We verify for each delay the configuration that achieves the upper bound,
according to the methodology given in Section~\ref{subsec:raa}
 (labeled ``Upper Bound''). We observe that, despite the high cost and the need for an optimal calculation of the end-to-end delay per connection
and the packet loss rate, the upper bound algorithm only slightly increases the aggregate throughput observed by the station respect to the \emph{min-max disconnection time} approach. The main reason behind this result is that the key parameter that affects the end-to-end TCP throughput is 
the disconnection time from the AP, which is taken into account in the \emph{min-max disconnection time} approach, rather than the extra-buffering time at the AP. 

\textbf{Case 2: VSTAs have different duty cycle and different delays}
We now consider a station connected to 3 APs, with the scheduler giving an output the set: $f_1=0.5$, $f_2=0.125$, $f_3=0.375$.
The corresponding number of slots given by the resource allocation algorithm are $g_1=4$, $g_2=1$, $g_3=3$.
We also suppose that $\mathrm{SlotTime}=12.5$\,ms.

In this example, we consider that the $\mathrm{VSTA}s$ experience different delays. Particularly we suppose that for
a given delay $x$ on $\mathrm{VSTA}_1$, the delay on $\mathrm{VSTA}_2$ is $x+20$\,ms and the delay on $\mathrm{VSTA}_3$ is $x+40$\,ms.
We then calculate the aggregated throughput summing the throughput achieved on the three $\mathrm{VSTA}s$ at a function of the delay $x$.

Without using the resource allocation algorithm, $\mathrm{VSTA}_1$ would be disconnected $50$\,ms, $\mathrm{VSTA}_2$ for $87.5$\,ms, and $\mathrm{VSTA}_3$
for 62.5 ms. The \emph{min-max disconnection time} algorithm increases the granularity of the AP assignment so that $\mathrm{VSTA}_1$ will be disconnected for $12.5$\,ms, $\mathrm{VSTA}_2$ (still) for $87.5$\,ms,
and $\mathrm{VSTA}_2$ for at most $37.5$\,ms. Note that the TCP throughput at $\mathrm{VSTA}_2$ can be improved only by reducing the $\mathrm{SlotTime}$,
since it uses just one slot per period and a resource allocation algorithm cannot contribute to improve $\mathrm{VSTA}_2$ throughput.

\begin{figure}[t]
\centering
\subfigure[Slot distribution.]{\includegraphics[height=3.8cm]{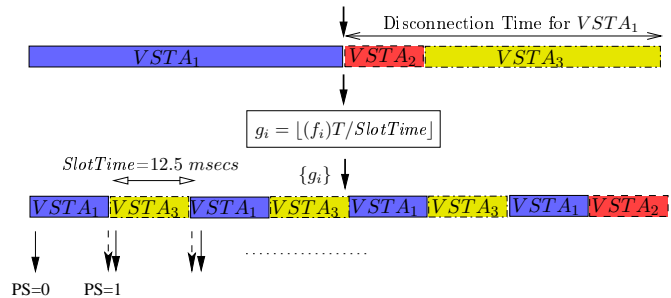}
\label{fig:MAC_TDMA}}
\hfil
\subfigure[Throughput improvement.]{\includegraphics[height=6cm]{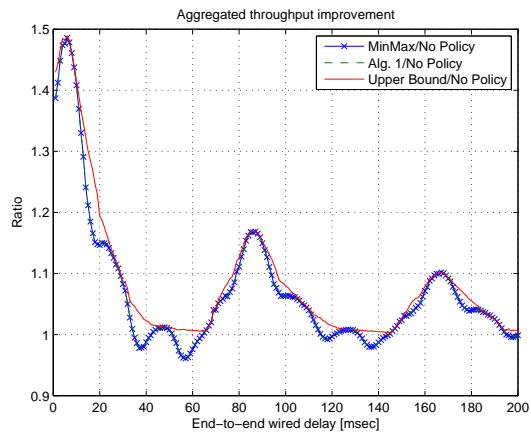}
\label{fig:throughput_with_Min-Distance_case2}}
\caption{Min-max disconnection time allocation algorithm. Case 2.}
\label{fig:xxx}
\end{figure}

In Fig~\ref{fig:throughput_with_Min-Distance_case2}, we observe that the ratio between the aggregated throughput obtained 
by the \emph{min-max disconnection time} algorithm over the one obtained without any algorithm is higher (up to 1.5 times) in most cases.
Particularly, the \emph{min-max disconnection time} allocation algorithm gets higher throughput corresponding to the scenarios where the delay added by the disconnection periods causes higher RTT. 
In general, the resource allocation algorithm can significantly improve the performance, since the VSTAs have different throughput demands. Our resource allocation algorithm tries to meet these demands by selecting the slot combination that minimizes the disconnection time. 

Besides, the ``Alg. 1/NoPolicy'' line in Fig~\ref{fig:throughput_with_Min-Distance_case2} is identical to the one achieved running the algorithm in (\ref{eq:brute_force}), that needs $280$ runs to analyze all the feasible solutions, respect to the $6$ runs needed by the \emph{min-max disconnection time} algorithm.
Finally, ``Upper Bound/NoPolicy'' line shows that the throughput improvements achieved with the optimum solution (upper bound algorithm)
is negligible. 

\begin{figure}[t]
\centering
\subfigure[Slot distribution.]{\includegraphics[height=3.8cm]{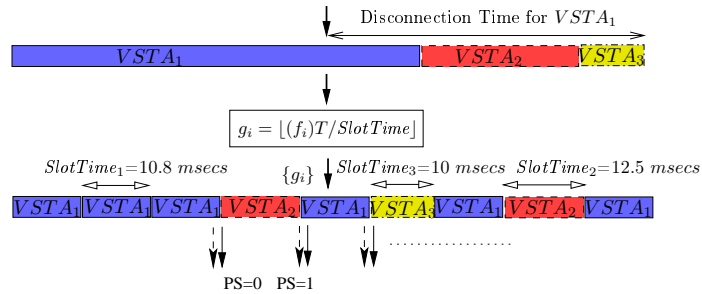}
\label{fig:MAC_TDMA_new}}
\hfil
\subfigure[Throughput improvement.]{\includegraphics[height=6cm]{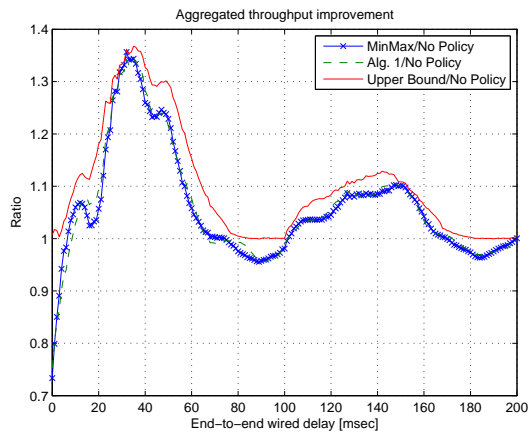}
\label{fig:throughput_with_Min-Distance_case3}}
\caption{Min-max disconnection time allocation algorithm. Case 3.}
\label{fig:xxx}
\end{figure}

\textbf{Case 3: Different slot size per VSTA}
We finally consider a scenario where the slot time length are different per each VSTA,
caused by a set of duty cycle equal to $f_1=0.65$, $f_2=0.25$, $f_3=0.10$.
The corresponding slot distribution is given in Fig.~\ref{fig:MAC_TDMA_new},
supposing $\mathrm{SlotTime}=10$\,ms. According to the slot allocation procedure 
defined in \ref{sub:slotted_operation}, the VSTAs use $\mathrm{SlotTime}_1=10.8$\,ms, 
$\mathrm{SlotTime}_2=12.5$\,ms and $\mathrm{SlotTime}_3=10$\,ms.

The results in Fig.~\ref{fig:throughput_with_Min-Distance_case3} show that
the \emph{min-max disconnection time} algorithm still achieves the best trade-off
between performance and computational cost. Particularly, we can see from ``Alg. 1/NoPolicy'' line,
that the throughput achieved by the \emph{min-max disconnection time} algorithm is very similar 
to the one defined in the equation \ref{eq:brute_force}.
Besides, in some scenario the \emph{min-max disconnection time} algorithm achieves a slightly higher throughput,
despite the cost of only $7$ runs compared to the $252$\,runs
needed in (\ref{eq:brute_force}).

\section{Conclusion}
\label{sec:conclusions}
The aggregation of the ADSL bandwidth via 802.11 wireless communication and multi-AP TDMA could dramatically increase the RTT observed by TCP flows. In this paper we studied this problem, both via extensive experiments with our prototype implementation
and via a simulator that accurately correlates the TCP RTT with the time spent by the wireless station on each AP. We presented a simple model that accurately
predicts the main effects caused by TDMA schemes on the observed TCP RTT, and we introduced a resource allocation algorithm that improves the aggregated throughput respect to state-of-the-art approaches with a complexity that grows linearly with the number of APs. Our solution does not require modifications to the rest of the network, and it can be applied to existing solutions that aggregate the AP backhaul bandwidth. Furthermore, we showed that its throughput performance is very close to the theoretical upper bound for a number of key scenarios. We believe that our approach will help to provide an efficient solution to aggregate multiple AP backhauls independently of the type of TCP traffic.

\bibliographystyle{IEEEtran}

\bibliography{bibliography}

\end{document}